\documentclass{svmult}
\usepackage{epsfig,graphicx} 
\usepackage{natbib,url}      
\bibpunct{(}{)}{;}{a}{}{,}   
\def\cite#1{\citealp{#1}}    
\def\authorindex#1{}
\def\figspath{.}  

\begin{document}\newcount\preprintheader\preprintheader=1



\title*{Spectroscopic Diagnostics of Polar Coronal Plumes}

\titlerunning{Polar Coronal Plumes from Space}

\author{K.~Wilhelm\inst{1}
        \and
        B.~N.~Dwivedi\inst{2, 1}
        \and
        W. Curdt\inst{1}
        }
\authorindex{Wilhelm,~K.}
\authorindex{Dwivedi,~B.~N.}
\authorindex{Curdt,~W.}

\institute{Max-Planck-Institut f\"ur Sonnensystemforschung, Katlenburg-Lindau, Germany.
           \and
           Department of Applied Physics, Institute of Technology, BHU, Varanasi, India.}

\maketitle

\setcounter{footnote}{0}  

\begin{abstract}
  Polar coronal plumes seen during solar eclipses can now be studied
  with space-borne telescopes and spectrometers.  We briefly discuss
  such observations from space with a view to understanding their
  plasma characteristics.  Using these observations, especially from
  SUMER/{\it SOHO}, but also from EUVI/{\it STEREO}, we deduce
  densities, temperatures, and abundance anomalies in plumes and
  inter-plume regions, and discuss their implications for better
  understanding of these structures in the Sun's atmosphere.
\end{abstract}

\section{Introduction}      \label{wilhelm-sec:introduction}

Polar coronal plumes are ray-like structures aligned along open
magnetic field lines in polar coronal holes. A total eclipse of the
Sun shows these rays in white light, depicting the magnetic
configuration of the Sun in a coronal hole. Many studies have been
carried out to relate these rays to the coronal magnetic field
inferred by current-free photospheric magnetic field
extrapolation. The coronal plumes and the inter-plume regions seem to
play a r\^ole in the acceleration mechanism of the fast solar
wind. They have been extensively observed from space across the
electromagnetic spectrum.  Investigations have been made to unravel
the appearance and disappearance of these plumes. The fact remains
that we know little about them, probably because we have no direct
knowledge of the coronal magnetic field. The identification of the
sources that produce coronal plumes and their contribution to the fast
solar wind is still a matter of investigation
\citep{
1997SoPhy..175...393, 1997ApJ..484...L75, 1998ApJ..500...1023,
2003ApJ..589...623, 2003ApJ..588...566, 2004A&A..416...749,
2008A&A..481...L61}.  To understand the processes of plume formation,
we need to know the physical conditions in plumes and the surrounding
inter-plume environment, such as electron densities, $n_{\rm e}$, and
electron temperatures, $T_{\rm e}$, the effective ion temperatures and
non-thermal motions, the plume cross-section relative to the size of
the coronal hole, and the plasma bulk speeds.

In this paper, we briefly discuss the observations of polar coronal
plumes from space with a view to understanding their plasma
characteristics.  Using these measurements, especially from SUMER/{\it
SOHO}, we deduce electron densities and temperatures as well as
abundance anomalies in plumes.  This will improve the understanding of
these structures in the Sun's atmosphere, which are the subject of an
International Team Study at ISSI, Bern\footnote{http://www.issibern.ch/teams/solarcoronal}.

\section{Spectroscopic Observations of Coronal Plumes}  \label{wilhelm-sec:observation}

\begin{figure}

  \centering
  \includegraphics[width=\textwidth]{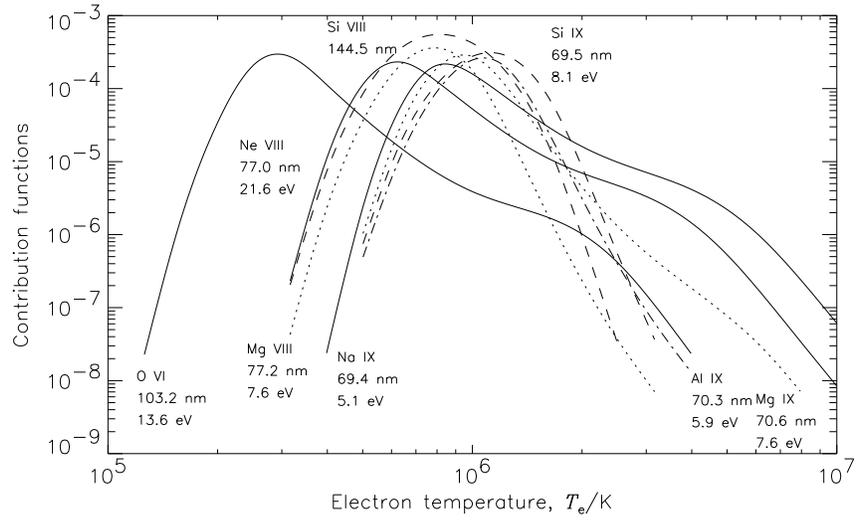}
  \caption[]{\label{wilhelm-fig:contrib}

Contribution functions of  the  observed  lines  and  the
FIP values of the corresponding elements, based on ionic fractions from
\citet{1998A&AS..133...403}. Neon and oxygen have high FIP values,
whereas the other elements have low values $< 10$ eV.}

\end{figure}

In the framework of a {\it Hinode/STEREO/SOHO} cooperation, observations of coronal
plumes in a coronal hole were performed in April 2007, using spectrographs and imagers
(cf., EUVI/{\it STEREO}) aboard these spacecraft \citep[cf.,][]{2008A&A..481...L61}.
SUMER performed a scan in the southern coronal hole of the Sun from
7 April 2007, 01:01 UTC to 8 April 2007, 12:19 UTC. Emission was observed from the
O\,{\sc vi}, Ne\,{\sc viii}, Mg\,{\sc viii}, Mg\,{\sc ix}, Si\,{\sc viii},
Si\,{\sc ix}, Al\,{\sc ix}, and Na\,{\sc ix} lines. The spectral lines were
recorded almost simultaneously at each location. Contribution functions and
the FIP (First Ionization Potential) values of the corresponding elements are shown in Fig.~\ref{wilhelm-fig:contrib}.
Density-dependent Si\,{\sc viii} and temperature-dependent
Mg\,{\sc ix} line ratios were observed to produce $n_{\rm e}$ and $T_{\rm e}$ maps.
Line-width studies allowed us to monitor the ion temperatures, which are much
higher than the electron temperatures.

\section{Results and Discussion}   \label{wilhelm-sec:discussion}

\begin{figure}

  \centering
  \includegraphics[width=\textwidth]{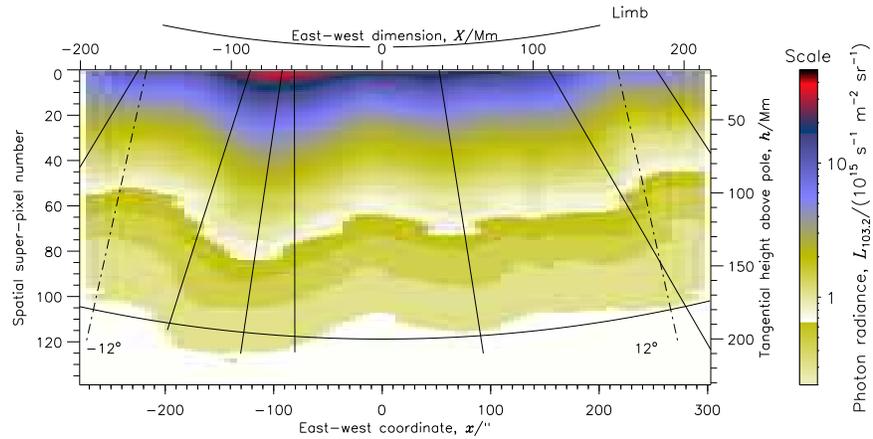}
  \caption[]{\label{wilhelm-fig:raster}

A large raster above the southern coronal hole was obtained in several VUV emission lines.
It took 36~h starting on 7~April 2007 at 01:01 UTC from West to East.
The photon radiance of  O\,{\sc vi} $\lambda 1032$ is shown here.
Radial dashed-dotted lines are shown at  $\pm 12^\circ$ off the pole.}

\end{figure}

Figure~\ref{wilhelm-fig:raster} shows a large raster above the
southern coronal hole, obtained in several VUV emission lines. All
maps are noisy above a height of 150~Mm.  The density and temperature
maps are, therefore, averaged over larger height ranges along the line
of sight. A detailed analysis of similar observations obtained
in 2005 has shown that the plume density is about five times higher
than that of the environment in this altitude range
\citep{2006A&A..455...697}.  The electron temperature, $T_{\rm e}$, in
plumes is lower than in interplume regions
\citep[cf.,][]{1998ApJ..500...1023}. This is confirmed by the Mg\,{\sc
ix} $T_{\rm e}$-sensitive line pair in the present data. The
insensitivity of the Si\,{\sc viii} ratio to scattered radiation is
discussed by \citet{1998ApJ..500...1023}.  It is caused by the lines
being barely visible on the disk \citep{2001A&A..375...591} so that
the stray-light is subtracted by the standard background correction
for coronal observations.

\begin{figure}

  \centering
  \includegraphics[width=\textwidth]{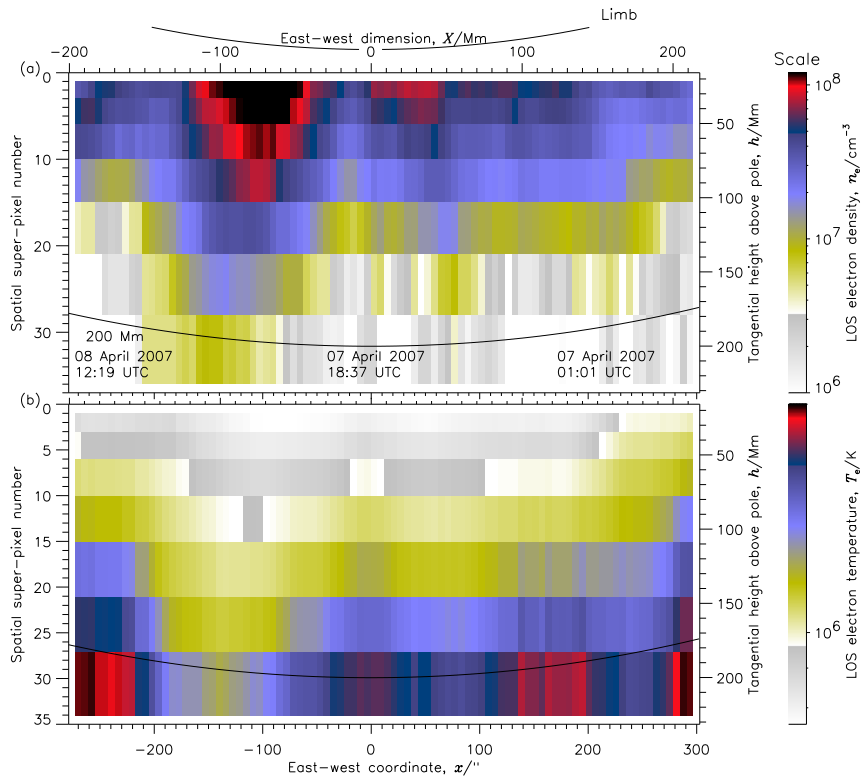}
  \caption[]{\label{wilhelm-fig:density}
(a) Electron density map;
(b) Electron temperature map (gray represents cooler plasma conditions).}
\end{figure}

Electron density and temperature maps are shown in
Figs.~\ref{wilhelm-fig:density}a and 3b, respectively, in which gray
represents cooler plasma conditions.  All radiance maps (except for
Al\,{\sc ix}) show the plume structures.  It is still to be
investigated why the Al\,{\sc ix} radiance map does not show plume
structure. The line ratio Ne\,{\sc viii}/Mg\,{\sc viii} can monitor
the abundance variations between high-FIP and low-FIP
elements. However, the different temperatures in plume and inter-plume
regions should be taken into account in view of the high-temperature
tail of the lithium-like Ne$^{7+}$. The contribution functions of
Ne\,{\sc viii} and Mg\,{\sc viii} overlap considerably in the
temperature range just below 1~MK. The contribution functions of
Ne\,{\sc viii} and Mg\,{\sc ix} are more similar at higher
temperatures than that of Ne\,{\sc viii} and Mg\,{\sc viii}. And yet,
the same signature is visible which indicates that a temperature
effect hardly plays a r\^ole. In particular, the Ne\,{\sc
viii}/Na\,{\sc ix} ratio shows that the abundance anomaly is real and
not a temperature effect. The estimated FIP bias for Ne and Mg is 1.5
to 2.  The search for Doppler shifts in the O\,{\sc vi} line did not
indicate any significant flows in plumes, although some of them were
directed out of the plane of the sky
\citep[cf.,][]{2008A&A..481...L61}. The super-radial expansion of plumes is evident
in the O\,{\sc vi} map.

\pagebreak
Investigating lines of high- and low-FIP elements in the SUMER data
set, we find:

\leftmargini=3ex
\begin{itemize} \itemsep=1ex

\item a temperature effect hardly plays a r\^ole,
      reconfirming the existence of abundance anomalies in plumes;

\item $T_{\rm e}$ is lower in plumes than in inter-plume regions in agreement with earlier findings;

\item plumes have higher densities than their environment;

\item no significant flows could be detected in coronal plumes  below 150~Mm.

\end{itemize}

Implications of deducing physical parameters, such as ion and electron
temperatures, densities, abundance anomalies, outflow velocities in
plumes and inter-plume regions are crucial to develop theoretical
models of these features and of high-speed solar wind. This will also
help in understanding the plume footpoint (e.g., XBPs), the
relationship between processes at the footpoint and the plume
characteristics, and to explore whether there is a relationship
between plumes and the fast solar wind. We plan to expand this work
and to communicate it as an article to the Astrophysical Journal.


\begin{small}

\end{small}

\begin{thebibliography}{4}

\expandafter\ifx\csname natexlab\endcsname\relax\def\natexlab#1{#1}\fi

\bibitem[Antonucci {et~al.}(2004)]{2004A&A..416...749}
Antonucci,~E., Dodero,~M.~A., Giordano, et~al. 2004, A\&A, 416, 749

\bibitem[Curdt {et~al.}(2001)]{2001A&A..375...591}
Curdt,~W., Brekke,~P., Feldman,~U., et~al. 2001, A\&A, 375, 591

\bibitem[Curdt {et~al.}(2008)]{2008A&A..481...L61}
Curdt,~W., Wilhelm,~K., Feng,~L., \& Kamio,~S. 2008, A\&A, 481, L61

\bibitem[DeForest {et~al.}(1997)]{1997SoPhy..175...393}
DeForest, ~C.~E., Hoeksema,~J.~T., Gurman,~J.~B., et~al. 1997, Sol. Phys., 175, 393

\bibitem[Gabriel {et~al.}(2003)]{2003ApJ..589...623}
Gabriel,~A.~H., Bely-Dubau,~F., \& Lemaire,~P. 2003, ApJ, 589, 623

\bibitem[Mazzotta {et~al.}(1998)]{1998A&AS..133...403}
Mazzotta,~P., Mazzitelli,~G., Colafrancesco,~S., \& Vittorio,~N. 1998, A\&AS, 133, 403

\bibitem[Teriaca {et~al.}(2003)]{2003ApJ..588...566}
Teriaca,~L., Poletto,~G., Romoli,~M., \& Biesecker,~D.~A. 2003, ApJ, 588, 566

\bibitem[Wang {et~al.}(1997)]{1997ApJ..484...L75}
Wang,~Y.-M., Sheeley,~N.~R.,~Jr., Dere,~K.~P., et~al. 1997, ApJ, 484, L75

\bibitem[Wilhelm(2006)]{2006A&A..455...697}
Wilhelm,~K. 2006, A\&A, 455, 697 

\bibitem[Wilhelm {et~al.}(1998)]{1998ApJ..500...1023}
Wilhelm,~K., Marsch,~E., Dwivedi,~B.~N., et~al. 1998, ApJ, 500, 1023

\end{thebibliography}
\end{document}